\documentstyle[11pt]{article}
\begin{document}

\begin{flushright}
hep-th/0306121
\end{flushright}

\begin{center}
{\Large\bf $C$-FIELD COSMOLOGY IN HIGHER DIMENSIONS}\\[20mm]
S.Chatterjee\footnote{E-mail: sujit@juphys.ernet.in} and
A.Banerjee \footnote{E-mail: asitb@cal3.vsnl.net.in}\\
{\em Relativity and Cosmology Research Centre,\\
Physics Department, Jadavpur University, Kolkata 700 053,India}
\end{center}

\vspace{0.5cm}
{\em PACS Nos.: 04.20.Cv, 98.80}
\vspace{0.5cm}

\pagestyle{myheadings}
\newcommand{\be}{\begin{equation}}
\newcommand{\ee}{\end{equation}}
\newcommand{\bea}{\begin{eqnarray}}
\newcommand{\eea}{\end{eqnarray}}
\begin{abstract}

Hoyle and Narlikar's $C$-field cosmology is extended in the framework of higher dimensional 
spacetime and a class of exact solutions is obtained. Adjusting the arbitrary constants of 
integration one can show that our model is amenable to the desirable property of dimensional 
reduction so that the universe ends up in an effective 4D one.Further with matter creation 
from the $C$-field the mass density steadies with time and the usual
bigbang singularity is avoided. An alternative mechanism is also suggested which seems to 
provide matter creation in the 4D spacetime although total matter in the 5D world 
remains conserved. Quintessence phenomenon and energy conditions are also discussed 
and it is found that in line with the physical requirements our model admits a
solution with a decelerating phase in the early era followed by an accelerated 
expansion later. Moreover, as the contribution from the $C$-field is made negligible a class of our 
solutions reduces to the previously known higher dimensional models in the framework of
Einstein's theory.

KEY WORDS: C-Field; Higher dimension;Singularity.
\end{abstract}

\section{INTRODUCTION}

For a long time the bigbang cosmology based on Einstein's field equations has come to be regarded 
as the only model that successfully explains the three important observations in astronomy, namely the
phenomenon of the expanding universe, primordial nucleosynthesis and the observed isotropy of the cosmic
microwave background radiation.However, the more sophisticated astronomical observations in the 
late eighties have revealed that the predictions of the Friedmann-Robertson-Walker type of models
do not always exactly meet our expectations as was believed earlier \cite{smoot}. Some puzzling results
regarding the redshifts from the extra galactic objects continue to evade theoretical
explanation from the  bigbang type of model,nor did the discovery of CMBR necessarily prove it 
to be a relic of the bigbang.In fact a nonrelic interpretation of the the CMBR is also
plausible\cite{jvn}.

So alternative theories are being proposed time to time -- the most wellknown being the Steady State 
theory by Bondi and Gold \cite{bg}. In this approach the universe does not have any singular beginning
nor an end on the cosmic time scale. Moreover the statistical properties of the large scale features
of the universe do not change. To account for the constancy of the mass density they have to envisage 
a very slow but continuous creation of matter going on in contrast to the one time infinite
and explosive creation at $t = 0$ of the standard model.But it suffers from the serious
disqualification that they do not give any physical justification in the form of any dynamical
theory for the phenomenon of the continuous creation of matter, however insignificant. Thus the
priciple of conservation of matter is clearly sacrificed in this formalism. To overcome this difficulty 
in particular and a host others Hoyle and Narlikar \cite{hn} in the sixties
 adopted a field theoretic approach introducing a massless and chargeless scalar field $C$ in
the Einstein-Hilbert action to account for the matter creation.In the $C$-field theory introduced
by Hoyle and Narlikar there is also no bigbang type of singularity as in the earlier steady
state theoy of Bondi and Gold.

In the present work we have thought it worthwhile to extend the pioneering work of Hoyle
and Narlikar in the framework of higher dimensional spacetime.Over the years the interests in higher
dimensional theories have stemmed in their attempts to unify gravity with other forces in nature
\cite{wesson}. The recent spurt in activities is also due to its new field of application in brane
cosmology\cite{horowitz}. Some of the important findings in the 5D C-field model of homogeneous
dust universe in contrast with the 4D model of HN may be summarised as follows:

\noindent
1. ~Unlike the 4D model of HN, in one case the dust density decreases continuously and finally
vanishes. In this specific model the rate of fall of the dust density vanishes at both extremes
$t= \pm \infty$.

\noindent
2. ~One of our models demonstrates an explicit example of dimensional reduction,evolving into
an effective 4D spacetime.

\noindent
3. ~There is atleast one situation where our higher dimensional $C$ field cosmology starts from the
 a singularity of infinite mass density and evolves with a decelaration followed by an
 acceleration at later stage reflecting the characteristic feature of the socalled 'quintessence'
 model. One must refer to the current ideas about an accelerated universe explained in terms of dark energy 
characterised by an equation of state $\omega = p/\rho$ being more negative than $- 1/3$ after
 inclusion of other forms of matter.

Our paper is organised as follows:

\noindent
In section 2 we have briefly introduced the original work of Hoyle and Narlikar. In section
3 an exact solution of HN theory in a 5D homogeneous spacetime  is found. The section 4
deals with the dynamical behaviour of our model where one of the cases exhibits the 
desirable property of dimensional reduction such that with time the universe ends
up as an effective 4D one. The section 5 is particularly interesting in the sense
that we here present a case where the cosmology manifestly has a bigbang type of singularity.
It emphasises the fact that mere inclusion of a $C$-field is no guarntee against the 
occurence of singular epoch.The section further deals with the {\it quintessence} phenomenon where
interestingly one of the models yields a decelerating universe in the early phase and 
later the gravitationally self repulsive $C$ field causes the expansion to accelerate. This
is in conformity with the present day observational status. The section 6 deals with the 
energy conditions in the framework of HN theory while in section 7 an alternative mechanism
is {\it naively} suggested where the matter creation in the 3D space apparently occurs as a
consequence of spontaneous compactification of the extra dimensions although the total matter in the 
higher dimensional sense is strictly conserved. We conclude in section 8.

\section{HOYLE-NARLIKAR THEORY}

As pointed out earlier Einstein's field equations are modified in this formalism through the
introduction of an external $C$ field such that (see ref. \cite{book} for more details)
\be
R_{ik} - \frac 1 2 g_{ik} R = - 8 \pi \left(^{m}T_{ik} + ^{c}T_{ik}\right) 
\ee
where the first term in the r.h.s. refers to the matter tensor of Einstein's theory and the second term is
that due to the $C$-field given by
\be
^{c}T_{ik} =  -f \left(C_{i}C_{k} - \frac 1 2 g_{ik} C^{\alpha} C_{\alpha}\right)
\ee
where $f > 0$ and $C_{i} = \frac{dC}{dx^i}$. 

Since $T^{00} < 0$ we see that the $C$ field has negative energy density producing repulsive gravitational
field driving the expansion of the universe. Thus the energy conservation law reduces to
\be
^{m}T^{ik}_{;ik} = - ^{c}T_{;ik} = f C^{i}C^{k}
\ee
i.e. matter creation through a nonzero l.h.s. is possible while conserving the overall energy and momentum.
The last equation is identical with
\be
m \frac{dx^{i}}{ds} g_{ik} - C_{k} = 0
\ee
which tells us that the  4-momentum of the created particle is compensated by 4-momentum of the $C$ field.
Clearly to achieve this balance the $C$ field must have negative energy. Further the $C$ field satisfies
the source equation $fC^{i}_{; i} = j^{i}_{; i}$  and  $j^{i} = \rho \frac{dx^i}{ds} = \rho v^i$
where $\rho$ is the homogeneous mass density.

\section{FIELD EQUATIONS AND ITS INTEGRALS}

We here discuss a spatially flat 5D homogeneous cosmological model with the topology
$M^1 \times R^3 \times S^1$ where $S^1$ is taken in the form of a circle such that
\be
ds^2  = dt^2 - R^{2}(t) \left(dr^2 + r^2 d\theta^2 + r^2 \sin^2 \theta d \phi^2 \right) - A^{2}(t) dy^2
\ee
where $R(t)$ is the scale factor for the 3D space and $A(t)$, that for the extra dimension.

\noindent
The independent field equations for our metric (5) and energy momentum tensor (2) are
\bea
3 \frac{\dot{R}^2}{R^2}  + 3\frac{\dot{R}}{R} \frac{\dot{A}}{A}  &=& 8 \pi \left(\rho -\frac{1}{2}
f \dot{C}^2 \right) \\
2 \frac{\ddot{R}}{R}  + \frac{\dot{R}^2}{R^2}  + 2 \frac{\dot{R}}{R}\frac{\dot{A}}{A} + 
\frac{\ddot{A}}{A}  &=& 4 \pi f \dot{C}^2 \\
3\frac{\ddot{R}}{R} + 3 \frac{\dot{R}^2}{R^2} &=& 4 \pi f \dot{C}^2
\eea
Following Hoyle and Narlikar we have also taken a zero pressure matter field in this work.

\noindent
From the Bianchi identity we further get
\be
\dot{\rho}  +  \left( 3 \frac{\dot{R}}{R} + \frac{\dot{A}}{A} \right)\rho = f \dot{C} 
\left[ \ddot{C} + \left
(3\frac{\dot{R}}{R} + \frac{\dot{A}}{A} \right)\dot{C} \right]
\ee
which when used in the source equation yields $\dot{C} = const. = 1$.

For economy of space we shall not give here the details of the intermediate mathematical analysis
to solve the field equations and mention important steps only. 
We substitute $\dot{R}= Y$ such that $\ddot{R} = \frac{dY}{dt} = \frac{dY}{dR}\frac{dR}{dt}
= Y Y'(R)$ where a prime overhead denotes differentiation with respect to $R$ and now $R$ becomes the
 new independent variable.
The equation (8) now reduces to 
\be
YY'+ \frac{Y^{2}}{R} - \frac{p^{2}R}{2}  = 0
\ee    
where $p^2 = 8{\pi}f/3$.
It yields a first integral as 
\be
Y^{2}R^2 = k + \frac{p^{2}R^4}{4}
\ee
where $k$ is an arbitrary constant of integration. From the last equation we finally get 
\be 
R^2 = {R_{0}}^{2} \cosh pt                                 
\ee
when $k < 0$.                                                                            
\be
R^2 = R_{0}^2 \sinh pt                                   
\ee
when $k > 0$.

\noindent
Moreover from the other two field equations (7)-(8) we obtain
\be
\ddot{A} + 2 \dot{A} \frac{\dot{R}} R  - A \left(\frac{\ddot{R}} R + 2 \frac{\dot{R}^2}{R^2}\right) = 0
\ee
By inspection we get a particular solution of the above equation as $A = R$. 

Following the standard method of solving this type of second order linear differential equation
we assume $A(t) = R(t)u(t)$ such that the last equation reduces to 
\be
R\ddot{u} + 4\dot{R}\dot{u} = 0
\ee
yielding successively after two quadratures
\be
\dot{u} = \frac{\beta}{R^4} = \frac{\beta}{{R_0}^{4} \cosh^{2}pt}
\ee
and 
\be
u = \frac{\beta}{R_0^4} (\gamma + \tanh pt )
\ee
where $\gamma$ is an arbitrary constant of integration.
So we finally get as general solution
\be
A = u R = \frac{a \cosh pt + b \sinh pt}{\sqrt{\cosh pt}}
\ee

Using the above values of the metric coefficients we get the expression for the mass density as
\be
\rho = \frac{f}{2} \frac{a \cosh^{2} pt + a \sinh^{2} pt + 2b \cosh pt \sinh pt}
{\cosh pt ( a \cosh pt + b \sinh pt)}
 = f \left[1 - \frac a {2 \cosh pt (a \cosh pt +  b \sinh pt)}\right]
\ee
From the expression of the energy density we note that $f > 0$.

\section{DYNAMICAL BEHAVIOUR}

Depending on the nature of the arbitrary constants of integration several interesting possibilities 
of the evolution present itself.

\noindent
{\bf Case I}~$(b = 0)$

\noindent
Our solutions reduce to
\be
R^2 = {R_0}^{2} \cosh pt, ~A^2 = a^2 \cosh pt ,~\rho = \frac{f (1 + \tanh^2{pt})} 2 ,~~~ 
\dot{\rho} = f \frac{\sinh pt}{\cosh^3 pt}
\ee
When ~$t\rightarrow -\infty$, $R^2 = {\frac1 2}{R_0}^{2}e^{-pt} \rightarrow \infty$
$A^2 = {\frac1 2}a^{2}e^{-pt} \rightarrow \infty$ whereas $\rho \rightarrow f$ and
$\dot{\rho} \rightarrow 0$.

\noindent
So in the range $-\infty < t \leq{0}$, the density decreases but reaches a steady value and $A^2$
also reduces to a steady value $a^2$, which is an arbitrary constant quantity. As $a$ may be chosen
as small as possible there is dimensional reduction but no big bang type of singularity at any stage.

\noindent
Moreover the 5D volume,$R^{3}A = R_{0}^{3}a\ cosh^2{pt}$ tends to infinity as
 $t\rightarrow\underline{+}\infty$
and to $R_{0}^{3}a$ (a finite value) as $t\rightarrow 0$. So the $5D$ volume decreases to a minimum
and then increases indefinitely. This model is rather puzzling in the sense that as the contraction
is followed by an expansion the mass density finally reaches a steady value -- even though both R and
A approach infinity, which is consistent with the steady state theory where matter creation via
 the creation field maintains the balance. The model gives rise to a very interesting probability
 that the higher dimension gains significance in future which follows a dimensional reduction without
 any inconsistency in the present day scenario of the universe. 
                                    
\noindent        
{\bf Case II}~$(a = 0)$

\noindent
The situation is distinctly different from what has been discussed in case I. Here

$R^2 = {R_0}^{2}\cosh pt,~ A^2 = b^2\frac{\sinh^2 pt}{\cosh pt}$    
and $\rho = f$ (constant always). The range $-\infty< t < 0$ is not valid here since the 5D volume
$R^3 A$ becomes  negative. But in the range $0 \leq t < \infty$, the 5D volume increases continuously
and indefinitely starting from zero. But the mass density remains constant throughout and the
behaviour mimics more a higher dimensional variant of the well known steady state theory of Bondi
and Gold where the matter density remains unchanged due to the hypothesis of continuous creation 
as mentioned in the introduction. This also follows from the divergence relation (9), which on integration
yields $\rho = f + \frac{k}{R^3 A}$. Comparision with (19) gives $k = \frac{f a R^{3}}{2}$.
For our case, $a = 0$ and hence $ k = 0$, giving   $ \rho = f $.  

\noindent
{\bf Case III}~$(b = - a)$

\noindent
In the context of higher dimemnsional theories this case is most interesting because with time the
extra dimensional scale factor shrinks exhibiting the desirable feature of dimensional 
reduction.We presume that as the extra space reduces to the planckian size some stabilising
mechanism (quantum gravity may be a possible candidate) would halt the shrinkage and the 5D cosmology 
becomes effectively four dimensional in nature.

\noindent
Here $R^2 = R_0^2 \cosh pt$, $A^2 = \frac{2a^2 e^{-2pt}}{e^{pt} + e^{-pt}}$~~~,
$\rho = \frac{f}{2} \left(1 - \tanh pt \right)$ and $\dot{\rho} = -f p/(2 \cosh^{2} pt).$

\noindent
Moreover as $t \rightarrow -\infty$, $R \rightarrow \infty$, $A \rightarrow \infty$, 
$\rho \rightarrow f$, $\dot{\rho} \rightarrow 0$

\noindent
as $t \rightarrow +\infty$, $R \rightarrow \infty$,
$A \rightarrow 0$, $\rho \rightarrow 0$, $\dot{\rho} \rightarrow 0$
 
\noindent
as $t \rightarrow 0$, $R \rightarrow R_0$, $A\rightarrow a$, $\rho\rightarrow f/2$, 
$\dot\rho\rightarrow\frac{-fp}{2}$.   
  
\noindent
So the density decreases continuously in the range $-\infty < t < +\infty$ starting from $\rho = f$  to
$\rho \rightarrow 0$ reaching steady state at two extremes $\dot{\rho} \rightarrow 0$.
In this case to find the exact asymptotic nature we see that as $t \rightarrow \infty$,
$R^2 \sim \frac{R_0^2}2 e^{pt}$ and $A^2 \sim 2 a^2 e^{-3pt}$.

\noindent
Hence dimensional reduction takes place and the 5D volume becomes $\frac{R_0^3 a^2}2$,a finite
magnitude. Thus there is no singularity throughout evolution.

\section{SINGULAR SOLUTION IN HN THEORY}

The creation field $C$ was introduced by Hoyle and Narlikar to avoid any big bang type of singularity
in cosmology sacrificing in the process the age old conservation principle. However we present here
a set of solutions in $C$ field theory in 5D spacetime with a big bang type singularity as follows:
\bea
R^2 &=&  {R_0}^2 \sinh pt \\
A &=& \frac{( a \sinh pt + b~ \cosh pt )}{\sqrt{\sinh pt}} \\
C &=& t \\
\rho &=&  \frac{f}{2}~ \frac{ a \sinh^2 pt + a \cosh^2 pt + 2 b \cosh pt \sinh pt }
{\sinh pt (a \sinh pt + b sinh pt)}.
\eea
This is obviously not a non singular model with everything blowing up at $t = 0$. However at
$t \rightarrow \infty$, the mass density becomes constant and $\dot{\rho}$ vanishes, a characteristic
of the steady state model. At this stage it may not be out of place to point out that this
type of singular solutions in $C$ fields cosmology is possibly unique in higher dimensional models only.
One can not obtain solutions of similar kind in 4D spacetime. So in the contex of higher dimensional
spacetime the $C$ field is not always effective in removing the perenial problem of singularity           
in cosmology.

\noindent
The above model in the framework of creation field cosmology in 5D shows an extraordinary character
in the behaviour of {\it deceleration parameter}. A little algebra shows that it is given by
\be
q = -\frac{R \dot{R}}{R^2} = \frac{1 - \sinh^{2} pt}{1 + \sinh^{2} pt}.
\ee
It is evident from (21) that the range $-\infty < t < 0$ is not valid for obvious reasons. However
the scale factor is real in the range $0 \leq t < \infty$. Initially $R \rightarrow 0$ and the
deceleration parameter is positive $(q\geq 0)$. But at $t = t_1$ with $\sinh^{2} pt_1 = 1$ we have $q = 0$,
whereas subsequently for $t > t_1$, that is in the range $t_1 < t < \infty$ the deceleration parameter
$q < 0$, which implies that the universe is accelerating consistent with the present day observations.
This result is quite interesting {\it vis a vis} the present day attempts to construct 
accelerating models as in quintessence \cite{cds}.

As the cosmology described in this section is somewhat similar to the big bang type it is tempting
to compare the set (21)-(24) with the well known multidimensional homogeneous models in the absence
of the $C$ field. Now as $f \rightarrow 0$, the contribution from the $C$ field becomes increasingly
insignificant. Since $4 \pi f = 3 p^2/2$, the equations (21)-(24), as $f$ tends to zero reduce to
$R \sim t^{1/2},~A \sim a t^{1/2} + b t^{-1/2}, ~\rho = \frac{a}{a p^2 t^2 + b p t}$.

This resembles our solution \cite{bcp} for a special case of homogeneous model when the 
inhomogeneity parameter is set to zero. If in addition we take $a = 0$, then $R \sim t^{1/2}$,
$A \sim t^{-1/2}$ and $\rho = 0$, which is the well known soluton of Chodos and DetWeiler \cite{cd} for
a 5D homogeneous empty universe. Relevant to point out that HN solution \cite{hn} are not amenable 
to similar reduction to FRW type of models when the creation field $C$ is switched off. 

\section{ENERGY CONDITIONS}

As discussed in the previous section Hoyle and Narlikar had to invoke an extraneous scalar field
with negative energy to get singularity free solutions.Following closely a recent work of Kolassis etal
\cite{santos}we shall discuss, very briefly, the weak, dominant and the strong energy conditions
in the context of $C$ field theory for our particular model.
With the energy momentum tensor given in section 2 we write for a zero pressure model in 4D
\be
T_{00} = ( \rho - \frac f 2 ) ,~~~ T_{11} = T_{22} = T_{33} = - \frac f 2
\ee
in the locally Minkowskian frame. Obviously the roots of the matrix equation
\be
|T_{ij} - r~g_{ij}| = diag~ [ (\rho - f/2 - r ),  (r - f/2 ),  (r - f/2), (r- f/2)~] 
\ee                    
give the eigenvalues $r$ of our energy momentum tensor as
~~~$r_{0} = (\rho - f/2 )$ ,  $r_1 =~ r_2 = r_3 =~ f/2$.

Skipping details( see ref. \cite{santos} ) the energy conditions for our model may be briefly summed up as:

\noindent
(a){\it Weak energy condition}
 
~~~ $r_0 \geq 0$~~ i.e., $\rho \geq f/2$ and ~~~ $(r_0 - r_i) \geq 0$~~~~~~i.e., $\rho \geq f$.

\noindent
(b){\it Dominant energy condition}

~~~ $r_0 \geq 0$~~ i.e., $\rho \geq f/2$~~~~$-r_0 \leq -r_i \leq r_0$~~~~~~i.e., $(-\rho + f/2)
\leq{-f/2} \leq (\rho - f/2)$.

\noindent
Obviously   $(\rho - f )\geq 0$ ~~~~i.e., $\rho > f$.

\noindent
(c) {\it Strong energy condition}

~~~ $(r_0 - \sum r_i) \geq 0$.~~~It follows that $\rho\geq{2f}~~ (r_0 - r_i) \geq 0$~~~ 
implying~~ $\rho \geq f$

\noindent
So all the conditions clubbed together lead to $\rho > 2f$ satisfying all the energy conditions.

\noindent
A little algebra shows that for a $(n+4)$ dimensional model the identical situation for 
strong energy condition leads to ~$( r_0 - \sum r_i) \geq 0$ ~~~ or $\rho\geq( 2f + \frac{nf} 2)$.
So apparently the extra dimensions put more stringent condition for a physically realistic energy
momentum tensor.

\section{AN ALTERNATIVE PROPOSAL}

As pointed out in the section 2 that to avoid cosmological singularity the hypothesis of matter creation
by Gold and Bondi is an adhoc assumption without any dynamical mechanism being offered to justify the 
process. To circumvent this difficulty HN introduced the so called {\it  creation field} without much of
any plausible physical justification. In both the cases the conservation principle is clearly violated.
In the event that the spacetime has indeed extra dimension it is not suffient to have proved that 
spontaneous compactification occurs. Rather the cosmological consequences of the disappearance of the
extra dimensions should also be studied.Here we briefly describe a scenario where matter may be created
in the 3D space as a result of dimensional reduction of the extra space but matter conservation is still
valid in the higher dimensional sense.

\noindent
From the Bianchi identity it follows that in the absence of an external $C$ field 
\be  
\rho R^{3} A = const = M(0)
\ee
where $M(0)$ is the total mass in the 5D world which can not but conserve. But for a 4D observer 
the effective 4D matter will be given by $\rho R^{3} = M_{4}$ such that $M_4(t) = M(0)~A^{-1}$.
Since for a physically realistic model the extra dimensions should shrink the above equation tells us
although the overall (4+1)D matter remains conserved there will be {\it matter leakage} from the
internal space onto the effective 4D world. So it offers a natural mechanism for matter creation
in 4D spacetime without the assumtion of an extraneous field.

\section{DISCUSSION}

We extend to higher dimension an earlier work of HN in $C$-field cosmology. Our work presents 
varied scenarios. Depending on the nature of the arbitrary constants the extra dimension 
either expand along with the 3D space or it shrinks with time with the cosmology ending up as an
effective 4D one.We here chose a topology $M^{1} \times R^{3} \times S^{1}$ but we believe that
most of the findings may be extended if we take a large number of extra spatial dimensions. An important 
result of our investigation is the appearence of singular solution in the the $C$ field theory. This has
probably no analogue in the 4D spcetime and it seemingly suggests that the mere presence of a $C$ field
is no guarntee against singular solutions in cosmology. Moreover this set reduces to the wellknown
higher dimensional soltions in generalised Einstein's equations when contribution from the $C$ field is
made insignificant.

We have also put forward an alternative mechanism for matter creation in 3D space without invoking the
existence of an extraneous creation field. This occurs as a consequence of dimensional reduction of the
extra space although total matter is strictly conserved in the higher dimensional sense. However the 
idea is too premature to come to any definite conclusion in this regard and hence we are very brief on 
this point.

Recently Hoyle,Narlikar and Burbidge \cite{book} modified their previous theory to Quasi steady
state cosmology(QSSC) where the cosmology has an oscillatory phase superposed on a steadily expanding
de Sitter type solution of field equation. As both QSSC and higher dimensional spacetime are both
particularlyrelevant in the early phase of cosmic evolution an extension of the ideas of QSSC in the
realm of multidimensional cosmology is urgently called for.

\vskip .2in 
\noindent 
{\bf ACKNOWLEDGEMENT}

\noindent
SC wishes to thank Third World Academy of Science, Trieste for travel support and Institute of
Theoretical Physics, Beijing for local hospitality, where part of the work is done.

\end{document}